\documentstyle[prd,aps]{revtex}
\begin{document}
\title{Exact Equations and Scaling Relations for $\bar f_0$-avalanche in
the Bak-Sneppen Evolution Model}
\author{W. Li$^{1,}$\thanks{Electronic address: liw@iopp.ccnu.edu.cn} 
    and X. Cai$^{1,2}$\thanks{Electronic address: xcai@wuhan.cngb.com} \\ 
\footnotesize \sl 
$^{1}$Institute of Particle Physics, Hua-zhong Normal University, 
Wuhan 430079, China\\
$^{2}$Institut f$\ddot{u}$r Theoretische Physik, FU Berlin, Arnimallee 14, 14195
Berlin, Germany}

\maketitle
\vskip 0.5cm
\begin{abstract}
Infinite hierarchy of exact equations are derived for the newly observed 
$\bar f_0$-avalanche in
the Bak-Sneppen model. By solving the first order exact equation, 
we find that the critical exponent $\gamma$, governing the divergence
of the average avalanche size, is exactly $1$ 
(for all dimensions), which has been confirmed by extensive simulations. 
Solution of the gap equation yields another universal result $\rho =1$
($\rho$ is the exponent of relaxation to attractor). 
Scaling relations are established among the critical exponents
($\gamma$, $\tau$, $D$, $\sigma$ and $\nu$) for $\bar f_0$-avalanche.
  
\vskip 0.5cm
\noindent
PACS number(s): 87.10.+e, 05.40.+j, 64.60.Lx 
\end{abstract} 

\twocolumn
In the Bak-Sneppen (BS) evolution model [1], random numbers, $f_i$, chosen from a flat
distribution between $0$ and $1$, $p(f)$, are assigned independently to each species located on
a $d$-dimensional lattice of linear size $L$. At each time step, the
extremal site, i.e., the species with the smallest random number, and its $2d$
nearest neighboring sites, are assigned $2d+1$ new random numbers also
chosen from $p(f)$. This updating continues indefinitely. After a long
transient process the system reaches a statistically stationary state where
the density of random numbers in the system is
uniform above $f_c$ (the self-organized threshold) and vanishes for $f < f_c$ .

Despite the fact that it is an oversimplification of real biological
process, BS model exhibits such common interesting features observed by
paleontologists [2,3] as punctuated equilibria, power-law probability
distributions of lifetimes of species and of the sizes of extinction events.
These behaviors suggest that the ecology of interacting species might have 
evolved to a self-organized critical state.

BS model displays spatial-temporal complexity, which also emerges from many
natural phenomena, such as fractals [4], $1/f$ noise [5], etc. This strongly
suggests that various complex behaviors may be attributed to a common
underlying mechanism. Authors in Ref. [6] suggest that the relation of these different
phenomena can be established on the basis of their unique models. It is even
proposed by them that spatial-temporal complexity 
comes out as the direct results of
avalanche dynamics in driven systems, and different complex phenomena are
related via scaling relations to the fractal properties of the avalanches.
It hence can be inferred that avalanche dynamics plays a key role in dealing
with complex systems, especially when one needs to know the macroscopic
features of the systems, since lingering on the inner structure of
individuals will not be helpful [7].

Avalanche is a kind of macroscopic phenomenon driven by local interactions.
The size of an avalanche may be extremely sensitive to the initial
configuration of the system, while the distribution of the sizes (spatial
and temporal) of avalanches, i.e. the "fingerprint" should be robust with
respect to the modifications, due to the universality of complexity and the
definition of self-organized criticality (SOC) [8]. 
In this sense, the extent that
we know about avalanche will determine to what extent we do a complex
system. Avalanche dynamics provides insight into complexity and enables one
to further investigate the system studied.       

Though avalanche dynamics may be a possible underlying mechanism of
complexity, the definitions of avalanches can be vastly different for
various complex systems, or for same sorts of systems, even for 
the same one.
In BTW model [9], an avalanche is intrigued by the adding of a grain or
several grains of sand into the system. The avalanche is considered over when
the heights of all the sites are less than the critical value, say, $4$. In
BS model [1,6], several types of avalanches, for instance, $f_0$-avalanche,
$G(s)$-avalanche, forward avalanche and backward avalanche, etc, are
presented. These different definitions of avalanches may show their unique
hierarchal structures, while they manifest the common fractal feature of the
complex system, that is, SOC. It can be inferred that various types of 
avalanches are equivalent in the sense that they imply complexity.

Since similar structures and common features evidently arise in different
types of avalanches, it is straightforward that various avalanches
differ each other only in the contexts from which one comprehends them.
As known, the major aim of avalanche study is to investigate the universal
rules possibly hidden behind the evolution of the systems or the models. Hence,
the means of understanding the avalanches appear crucial. Better 
ways may enable one to know more about the system or the model and hence
to have better comprehension of the features corresponding to complexity.          
From this point of view, when studying avalanches one should try to 
choose the easier ways instead of the more difficult ones.
 
The evolution of the highly sophiscated BS model shows a hierarchal structure
specified by avalanches, which correspond to sequential mutations below
certain threshold. It has been noted [10] that in BS model an avalanche is
initiated when the fitness of the globally extremal site 
(the species with the least random
number) is larger than the self-organized threshold. That is, the triggering
event of an avalanche is directly related to the
fitness, the feature of individuals. In other words, the avalanche is directly associated with the
feature of individuals instead of general features of the ecosystem as a
whole. Is it feasible that the avalanches are directly intrigued by the
global feature of the whole system?  Can such global feature be expressed in
terms of the corresponding quantity?  If such quantity found and such
avalanches observed, may the new avalanches provide a new and easier way in
investigating properties of the model?

One of our previous works [10] presents such a different hierarchy of
avalanches ($\bar f_0$-avalanche) for BS model. We defined a global quantity, 
$\bar f$,
which denotes the average fitness of the system. The new type of avalanches are
directly related to $\bar f$. In this paper, 
we present a master equation for the hierarchal
structure of $\bar f_0$-avalanches. It prescribes the cascade process of smaller
avalanches merging into bigger avalanches when the critical parameter 
$\bar f_0$ is
changed. An infinite series of exact equations can
be derived from this master equation . The first order exact equation, together with an scaling ansatz
of the average sizes of avalanches, shows the exact result of $\gamma$, the
critical exponent governing self-organization, to be universally $1$ for all
dimensional BS models, which has been confirmed by extensive simulations of the
model. We also establish scaling relations related to some critical exponents
for $\bar f_0$-avalanche and make predictions on the values of some exponents.

The quantity $\bar f$ is a global
one of the ecosystem and can be expected to involve some general information 
about the whole system. It may represent the average population or living
capability of the whole species system. Larger $\bar f$ shows that the average
population is immense or the average living capability is great, and vice
versa. $\bar f$ is defined as

\begin{equation}
\bar f=\frac {1} {L^d} \sum \limits_{i=1}^{L^d} f_i ,
\end{equation}      

\noindent 
where $f_i$ is the fitness of the $i$th species of a system consists of
$L^d$ species. Let BS model start to evolve. At each time step of the evolution,
apart from the random numbers of the globally extremal site and its $2d$
nearest neighboring sites, the signal $\bar f$ is also tracked. Initially,
$\bar f$ tends to increase step-wisely. As the evolution continues further,
$\bar f$ approaches a critical value $\bar f_c$ and remain statistically
stable around $\bar f_c$. The plot of $\bar f$ versus time step $s$ shows that
the increasing signals of $\bar f$ follow a Devil's staircase [8], which
implies that punctuated equilibrium emerges. Denote $F(s)$ the gap of the punctuated equilibrium. Actually, 
$F(s)$ tracks the peaks in $\bar f$. After some careful derivation one can
write down an exact gap equation [6,10]

\begin{equation}
\frac {dF(s)} {ds}=\frac {\bar f_c- F(s)} {L^d \langle S \rangle
_{F(s)} } ,
\end{equation}

\noindent
where $\langle S \rangle_{F(s)}$ denotes the average size of avalanches
occurred during the gap $F(s)$ when $\bar f < F(s)$. This exact gap equation
will be exactly solved in this paper.

Signals $\bar f(s)$ play important roles in defining 
$\bar f_0$-avalanche. For any value of the auxiliary parameter $\bar f_0$ ($0.5 <
\bar f_0 <1.0$), an $\bar f_0$-avalanche of size $S$ is defined as a sequence
of $S-1$ successive events when $\bar f(s) < \bar f_0 $ confined between two
events when $\bar f(s) > \bar f_0$. This definition ensures that the
mutation events during an avalanche are spatially and temporally correlated.
It can also guarantee the hierarchal structure of the avalanches: larger
avalanches consists of smaller ones. As $\bar f_0$ is raised, smaller
avalanches gather together and form bigger ones. The statistics of $\bar
f_0$ will inevitably have a cutoff if $\bar f_0$ is not chosen to be $\bar
f_c$. This will not affect the size distribution 
provided that $\bar f_0$ approaches $\bar f_c$. Extensive simulations show that
exponents $\tau$ of $\bar f_0$-avalanche size distribution are 1.800 and
1.725 for 1D and 2D BS models respectively, amazingly  different from the
counterparts of $f_0$-avalanche, 1.07 and 1.245 [6]. This strengthens the
speculation that $\bar f_0$-avalanche is a different type of avalanche,
distinguished from any types of avalanches found previously.

Denote $P(S, \bar f_0)$ the probability of acquiring a $\bar f_0$-avalanche
of size $S$. The signals $\bar f(s)$ ($\bar f_0 < \bar f(s) < \bar f_0 +d
\bar f_0$) will stop the $\bar f_0$-avalanches and not ($\bar f_0 +d \bar
f_0$)-avalanches. That is , as $\bar f_0$ is raised by an infinitesimal
amount $d \bar f_0$ some of $\bar f_0$-avalanches merge together to form
bigger ($\bar f_0 +d \bar f_0$)-avalanches. This exhibits a hierarchal
structure of $\bar f_0$-avalanches and will be prescribed by the below exact
master equation. In some sense, the master equation reflects the ''flow'' of
probability of avalanche size distribution with respect to the change in 
$\bar f_0$.

Simulations show that $\bar f$ approaches $\bar f_c$ and remain statistically 
stable in the critical state. This feature is greatly different from the
feature of $f_{\rm min}$ (fitness of globally extremal site), which can
vary between 0 and 1. While $\bar f$ in the critical state fluctuates
slightly around $\bar f_c$. Therefore, the $\bar f_0$-avalanches will have
no good statistics if $\bar f_0$ is chosen as the value far less than $\bar f_c$,
since there only exists smaller avalanches in the model. To acquire a better
and reasonable distribution of $\bar f_0$-avalanches sizes, one should
choose the value of $\bar f_0$ under the condition $\bar f_0 \rightarrow
\bar f_c$. It should be emphasized that the master equation listed below is
valid also for $\bar f_0 \rightarrow \bar f_c$.     

Both theoretical analysis and extensive simulations suggest that the signals
$\bar f(s)$ which terminate $\bar f_0$-avalanches are uncorrelated and
evenly distributed between $(\bar f_0, \bar f_c)$ provided that $\bar f_0
\rightarrow \bar f_c$. The direct consequence of this observation is that
the probability of an $\bar f_0$-avalanche merging to 
$\bar f_0+d \bar f_0$-avalanche
is prescribed by $\frac {d \bar f_0} {\bar f_c - \bar f_0}$. It is important
to note that any two subsequent avalanches are mutually independent for the
following arguments to be true. In other words, the probability distribution
of $\bar f_0$-avalanches, initiated immediately after the termination of an
$\bar f_0$-avalanche of size $S$ is independent of $S$. This is true because
in BS model the dynamics within an $\bar f_0$-avalanche is completely
independent of the particular value of the signals $\bar f(s) > \bar f_0$ in
the background that were left by the previous avalanches.

Here present the master equation. As $\bar f_0$ is raised by an
infinitesimal amount $d \bar f_0$, the probability "flowing" out of the
size distribution of $\bar f_0$-avalanches is given by $P(S,\bar f_0) \frac {d \bar f_0} {\bar f_c-
\bar f_0}$, while the probability "flowing" into is given by
$\sum_{S_1=1}^{S-1} \frac {P(S_1,\bar f_0)} {\bar f_c-\bar f_0} P(S-S_1,\bar
f_0)$. Let $\bar
f_0 \rightarrow \bar f_c$ and $d \bar f_0 \rightarrow 0$, one can write down
the master equation as

\begin{eqnarray} \nonumber
(\bar f_c-\bar f_0) \frac {\partial P(S,\bar f_0)} {\partial \bar f_0}
&=&-P(S,\bar f_0) \\
&&+\sum \limits_{S_1=1}^{S-1} P(S_1,\bar f_0) P(S-S_1,\bar
f_0) .
\end{eqnarray} 

\noindent
The first term on the right hand of the equation expresses the loss of
avalanches of size $S$ due to the merging with the subsequent one, while the
second one describes the gain in $P(S,\bar f_0)$ due to merging of
avalanches of size $S_1$ with avalanches of size $S-S_1$.

In order to investigate the exact master equation it is convenient to make
some variable changes. Define $h=-\ln (\bar f_c- \bar f_0)$. Therefore, $\bar
f_0=\bar f_c$ corresponds to $h=+\infty$. Since in the master equation $\bar f_0$
is chosen to be close to $\bar f_c$, $h$ varies from a very large number to
$+\infty$. Due to the variable change the variable $h$ is chosen from the
distribution $P(h)=e^{-h}$, which seems to be more "natural". In the
following part we will use the new variable $h$ instead of $\bar f_0$. The
master equation can be rewritten, in terms of $h$, as

\begin{equation}
\frac {\partial P(S,h)} {\partial h}=-P(S,h)+\sum \limits_{S_1=1}^{S-1}
P(S,h) P(S-S_1,h) .
\end{equation}

Making Laplace transformation of Eq.(4), after some calculation, one
obtains 

\begin{equation}
\frac {\partial \ln (1-p(\beta,h))} {\partial h}=p(\beta,h) ,
\end{equation}

\noindent
where $p(\beta,h)=\sum \limits_{S=1}^{\infty} P(S,h)e^{-\beta S}$. This
exact  equation is the key one of this work. Many interesting physical
features can be derived from it. As $h < +\infty$ avalanches size will
have a cutoff. The normalization of $P(S,h)$ can be expressed as
$p(0,h)=\sum \limits_{S=1}^{\infty} P(S,h)=1$. Expanding both sides of 
Eq. (5) as Tylor
series throughout a neighborhood of the point $\beta=0$, one can immediately obtain

\begin{eqnarray} \nonumber
&\frac {\partial} {\partial h} [1-\langle S \rangle_h \beta+ \frac {1} {2!}
\langle S^2 \rangle_h \beta^2-\frac {1} {3!} \langle S^3 \rangle_h
\beta^3+...]=& \\ \nonumber
&[\langle S \rangle_h \beta- \frac {1} {2!}
\langle S^2 \rangle_h \beta^2+\frac {1} {3!} \langle S^3 \rangle_h
\beta^3+...] \times&\\
& [-1+\langle S \rangle_h \beta- \frac {1} {2!}
\langle S^2 \rangle_h \beta^2+\frac {1} {3!} \langle S^3 \rangle_h
\beta^3+...]& .
\end{eqnarray}

\noindent
Since the equation (6) holds for arbitrary $\beta$, comparing the
coefficients of different powers of $\beta$ in the above Taylor series gives
an infinite series of exact equations. Comparison of the coefficients of
$\beta^1$ results in

\begin{equation}
\frac {\partial \ln \langle S \rangle_h} {\partial h} =1 .
\end{equation}

\noindent
Eq. (7) is extremely interesting. Changing variable $h$ back into $\bar f_0$, one
can obtain the "gamma" equation [6,11]

\begin{equation}
\frac {d \ln \langle S \rangle_{\bar f_0}} {d \bar f_0}= \frac {1} {\bar
f_c-\bar f_0} . 
\end{equation}

\noindent
Inserting the scaling ansatz [6] $\langle S \rangle_{\bar f_0} \sim (\bar
f_c- \bar f_0)^{-\gamma}$ into Eq. (8), one immediately obtain an interesting
result

\begin{equation}
\gamma=1
\end{equation} .

\noindent
It should be noted that $\gamma=1$ is universal, that is, independent of the
dimension. The value of $\gamma$ for $\bar f_0$-avalanches is different 
from those for $f_0$-avalanche found in
Ref. [6], which are 2.70 and 1.70 for 1D and 2D BS models respectively.
Extensive simulations show $\gamma=0.99 \pm 0.01$ and $\gamma=0.98 \pm
0.01$ for 1D and 2D BS models respectively. Fig. (1) shows our simulation
results, which confirms the universal result $\gamma=1$. 

Higher powers of $\beta$ gives new exact equations. Here present the first
two

\begin{equation}
\frac {\partial} {\partial h} (\frac {\langle S^2 \rangle_h} {\langle S
\rangle_h})=2 \langle S \rangle ;
\end{equation}

\begin{equation}
\frac {\partial} {\partial h} (\frac {\langle S^3 \rangle_h} {3 \langle S
\rangle_h}-\frac {\langle S^2 \rangle_{h}^{2}} {2 \langle S
\rangle_{h}^{2}})
=\langle S^2 \rangle_h .
\end{equation}

Next present the solution of the exact gap equation for $\bar
f_0$-avalanches. Inserting the scaling relation $\langle S \rangle_{F(s)} \sim
(\bar f_c-F(s))^{-1}$ into the equation and integrating, one obtains 

\begin{equation}
\Delta \bar f(s)=\bar f_c-F(s) \sim (\frac {s} {L^d})^{-\rho}=(\frac {s} {L^d})^{-1} ,
\end{equation}

\noindent
where $\rho$ is the exponent of relaxation  to attractor [6]. Thus, we obtain
$\rho =1$. Interestingly, $\rho$ is also a universal exponent for all
dimensional BS models. It shows that the critical point ($\Delta \bar f=0$)
is approached algebraically with an exponent $-1$.

Up to now, we have obtained some exponents of corresponding physical
properties of $\bar f_0$-avalanches: $\tau$, avalanche size distribution
[10], $D$, avalanche dimension [10], $\gamma$, average avalanche size [10],
and $\rho$, relaxation to attractor [6]. Recall another two exponents [6]:
$\nu$, $\sigma$, which are defined as $r_{co} \sim (\bar f_c-\bar f_0)^{-\nu}$
and $S_{co}=(\bar f_c-\bar f_0)^{- \frac {1} {\sigma}}$ respectively. Here
$r_{co}$ and $S_{co}$ are referred to as the cut-off of the spatial extent
of avalanches (due to the limit system size ) 
and that of the avalanche size (due to the fact that $\bar f_0$ is not chosen as 
$\bar f_c$) respectively. It is natural to establish some scaling relations
of these exponents for $\bar f_0$-avalanches similar to those found in Ref.
[6,12] for $f_0$-avalanches. Nevertheless, these two types of avalanches
manifest similar fractal properties. Hence some common features should be
shared by them. Integrating of the  equation $\langle S \rangle=
\int_{1}^{(\bar f_c-\bar f_0)^{-\frac {1} {\sigma}}} 
S P(S,\bar f_0)dS$
and the scaling $\langle S \rangle \sim (\bar f_c- \bar f_0)^{-1}$ result in 

\begin{equation}
\gamma=\frac {2-\tau} {\sigma}=1 .
\end{equation}

\noindent
Due to the compactness [6] of avalanches, we have $S_{co} \sim
r_{co}^{D}=(\Delta f)^{- \nu D}$, thus

\begin{equation}
\nu=\frac {1} {\sigma D}=\frac {1} {(2-\tau) D} .
\end{equation}

\noindent
Eqs. (11)-(12) establish scaling relations among the critical exponents, and
they imply that the self-organization time to reach the critical state is
independent of the initial configuration of the system. A system of size $L$
reaches the stationary state when $[\Delta f(s)]^{-\nu} \sim L$. It can be
inferred from Eqs. (11)-(12) that, if one chooses $\tau$ and $D$ as two
independent exponents other exponents can be expressed in terms of them.
Among the six exponents mentioned above, $\tau$ and $D$ can be numerically
measured [10], $\gamma$ and $\rho$ can be analytically obtained, while $\nu$
and $\sigma$ are difficult to explore despite some methods measuring the
corresponding exponents for $f_0$-avalanches were introduced in
Ref. [13]. Therefore, we can rely on the scaling
relations and values of the exponents obtained to predict the values of
$\nu$ and $\sigma$. We predict $\sigma=0.2$ (1D) and 0.275 (2D),
$\nu=2.04$ (1D) and 1.17 (2D).

Comparing $\bar f_0$-avalanche with $f_0$-avalanche we find that the former
is more readily to be treated. Two critical exponents can be analytically
obtained and are found to be universal for all dimensional BS models.
Furthermore, the infinite hierarchy of exact equations and the exact gap
equations, together with their solutions, provide exclusive investigation of
the new type of avalanches. Another asset of $\bar f_0$-avalanche is that it
involves some information concerning the whole system. It can be concluded
that $\bar f_0$-avalanche does enable us to comprehend the complex system
from an effective and different context. The weak point of this avalanche 
is that it loses some knowledge of individuals. It is still unknown how these
individual features will matter. It is worthwhile to investigate the
avalanche dynamics further in the future.   

This work was supported in part by NSFC in China and Hubei-NSF. 
X.C. thanks T. Meng for hospitalities 
during his visit in Berlin.

\vskip 0.5cm
\begin{center}
\bf {Figure Captions}
\end{center}

\vskip 0.2cm

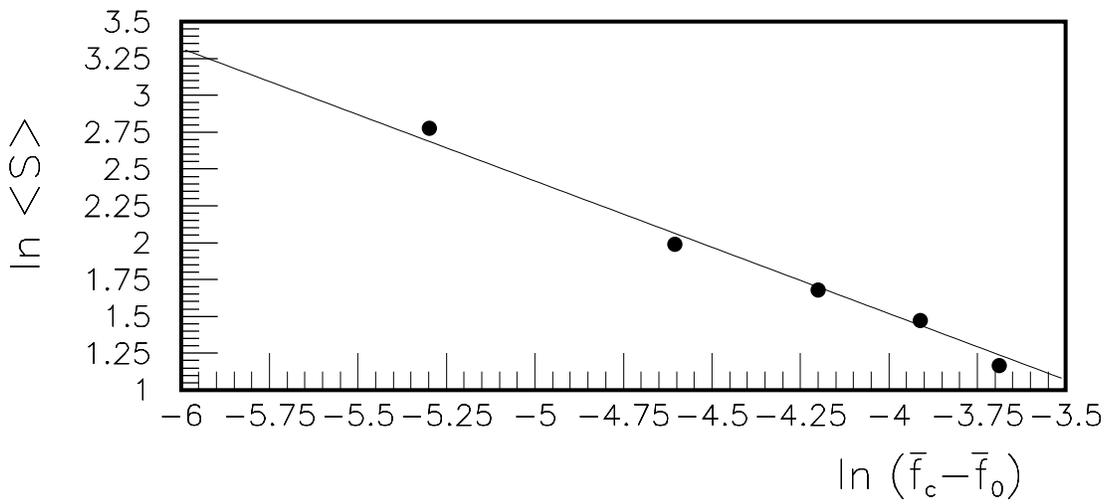
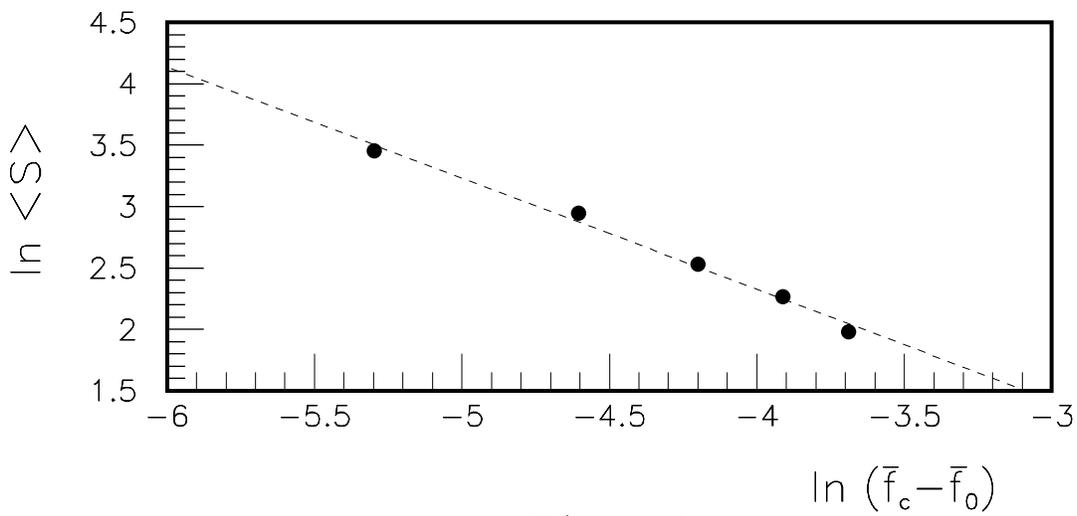
Fig. 1:  The average size of avalanches 
	 $\langle S \rangle$ vs $(\bar f_c-\bar
         f_0)$ for (a) 1D and (b) 2D Bak-Sneppen 
	 evolution models. The asymptotic slope yields
         $\gamma=0.99 \pm 0.01$ and $0.98 \pm 0.01$ respectively. 
\end{document}